Astrophysics

# THE SLOW DEATH OF MOST GALAXIES

**For most galaxies, the shutdown of star formation was a slow process that took four billion years. An analysis of thousands of galaxies suggests that 'strangulation' by their environment was the most likely cause. See Letter P. 192**

Andrea Cattaneo

In humans, death by strangulation is a slow process that takes about four minutes. During this time, the victim uses up oxygen in the lungs but keeps producing carbon dioxide, which remains trapped in the body. On page 192 of this issue, Peng *et al.*[1] present evidence of an analogous slow 'strangulation' process that ends the formation of stars in many galaxies by disrupting the supply of gas that accretes onto those galaxies from the environment. Instead of building up $CO_2$, the strangled galaxies accumulate metals produced by massive stars.

Ninety years ago, Edwin Hubble classified galaxies into three morphological types: spirals, lenticulars and ellipticals. In spirals, stars form a disc and turn around in circles like the horses of a cosmic merry-go-round. Ellipticals are the wrecks of galaxy crashes, in which stars move chaotically in all directions. Lenticulars form an intermediate type between the two. Most spirals are blue because they contain young stars, but elliptical and lenticular galaxies contain little or no cold gas to make new stars, and so only old red stars are left. Historically, astronomers have been more interested in the morphologies of galaxies than in their colour. Attention switched to colour after the Sloan Digital Sky Survey (SDSS) measured the spectra of hundreds of thousands of galaxies.

The SDSS demonstrated that blue (star-forming) galaxies and red (passive) galaxies form distinct populations[2]. Since then, various hypotheses have been put forward to explain what causes galaxies to transition from one type to the other. Most of them revolve around two ideas. The first is that the gas in ellipticals and their surroundings is too hot to make stars and does not cool efficiently. The second is that the gas that could cool and make stars is kept hot or blown away by phenomena linked to the growth of supermassive black holes, found at the centres of all ellipticals[3]. The main motivation for considering such scenarios comes from computer simulations of the formation of ellipticals in galaxy mergers. These simulations need a mechanism that gets rid of gas to avoid forming blue cores[4], which are rare in real ellipticals.

Peng *et al*. present evidence that the formation of stars in most passive galaxies ended through a slow strangulation process. The authors compared the metal content of the stars of approximately 23,000 passive galaxies from the SDSS with that of a control sample of about 4,000 star-forming galaxies, also from the SDSS. They discovered that the former is systematically larger than the latter, at least for galaxies that have stellar masses up to 100 billion times the mass of the Sun, the limit mass $M_*$ above which galaxies become scarce. This constitutes evidence for galactic 'suffocation', in the same way that high levels of $CO_2$ in the blood of a corpse suggest suffocation.

From the difference in the metal content of the stars of passive and star-forming galaxies, Peng *et al*. inferred a delay of 4 billion years (or 2 billion years for galaxies close to the limit mass) between the time when gas stopped being supplied and the time when star formation ended. This delay is consistent with the mean age difference between passive and star-forming galaxies (about 4 billion years at all masses).

As any forensic scientist will tell you, suffocation does not imply strangulation. But the difference in metal content is higher for galaxies in groups and clusters than it is for isolated ones, suggesting that crowded environments strangle galaxies by disrupting the accretion of gas onto them. This disruption might occur either through ram pressure (the pressure exerted on a body moving through a fluid) or through tidal forces.

The slow shutdown of star formation inferred by Peng *et al*. from observations of galaxies with masses lower or equal to $M_*$ contrasts with the fast shutdown behaviour of much larger galaxies, such as giant ellipticals. The stars of giant ellipticals have low iron content because they were made on a short timespan (less than 0.3 billion years for a galaxy of mass greater than $3M^*$) — that is, there wasn't enough time for many type Ia supernovae, the source of iron, to explode. Because there are far more red galaxies below $M_*$ than there are above, Peng *et al*. are correct to argue that strangulation is the main mechanism for star-formation shutdown. But the different chemical properties of low- and high-mass red galaxies imply that they must have formed through different routes. Morphological differences back this interpretation: red galaxies with masses above $M_{\text{star}}/3$ are all ellipticals or lenticulars; but below $M_{\text{star}}/3$, 40% of them are red discs[5], as expected if strangulation has occurred.

There is also a difference between ellipticals and lenticulars. Most elliptical galaxies in the stellar mass range 1 to $2M^*$ were already in place at redshift between 2 and 3, a period when the Universe was a fraction of its present age. Lenticulars appeared more gradually, replacing a pre-existing population of spiral and irregular galaxies (Marc Huertas-Company, private communication). The ratio of ellipticals to lenticulars is approximately 3:5 for galaxies around $1.5M_*$, but lenticulars should predominate in the mass range explored by Peng *et al*. Comparing the evolution of galaxies in star formation rate with their evolution in size provides further evidence for the presence of two evolutionary tracks[6].

Cosmological models[7] of the formation and evolution of galaxies predict two mechanisms for shutting down star formation. In these models, galaxies more massive than $1.5M_{star}$ form at the centres of groups and clusters. They grow extremely rapidly until, at about a redshift of 3 (when the Universe was a quarter of its current size), they attain the critical mass at which infalling gas is effectively shock-heated. Some violent phenomenon then quenches star formation. However, the models also predict that galaxies below $0.6M_{star}$ become red much later (at redshift less than 0.5) and far more gradually, in most cases because they stop accreting gas after becoming part of a group or a cluster. Thanks to the work of Peng *et al.*, this second theoretical prediction is now an observational fact.

Andrea Cattaneo is in the Observatoire de Paris, GEPI, 61 avenue del'Observatoire, 75014 Paris, France
e-mail: andrea.cattaneo@obspm.fr

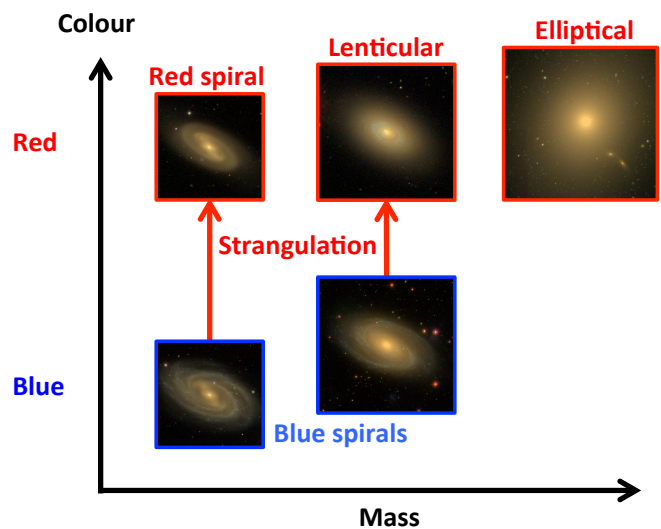

*A schematic picture of the formation of red galaxies. Slow strangulation of gas accretion shuts down star formation and causes blue spirals to evolve into red spirals or lenticulars if the disc fades so much that the spiral arms are no longer visible. Giant ellipticals have a different formation history, characterised by rapid shutdown of star formation, but these galaxies are much less numerous than red spirals.*